\documentclass[prc,aps,superscriptaddress,twocolumn,showpacs,nofootinbib]{revtex4}
\usepackage{graphicx}
\usepackage{amsmath}
\usepackage{amssymb}
\usepackage{dcolumn}
\usepackage{bm}

\newcommand{\nn}{\nonumber}
\renewcommand{\vec}[1]{{\mathbf #1}}
\newcommand{\vsigma}{\boldsymbol{\mathbf\sigma}}
\newcommand{\vnabla}{\boldsymbol{\mathbf\nabla}}

\newcommand{\nuc}[2]{$^{#1}${#2}}

\begin{document}

\title{New parameterization of Skyrme's interaction for
       regularized multi-reference energy density functional calculations}

\author{K. Washiyama}
\altaffiliation[Present address: ]{RIKEN Nishina Center, Wako 351-0198, Japan}
\affiliation{PNTPM, Universit\'e Libre de Bruxelles, 1050 Bruxelles, Belgium}

\author{K. Bennaceur}
\affiliation{Universit\'e de Lyon, F-69003 Lyon, France; Institut de Physique
Nucl\'eaire de Lyon, CNRS/IN2P3, Universit\'e de Lyon 1, F-69622 Villeurbanne,
France}

\author{B. Avez}
\affiliation{Universit\'e Bordeaux,
             Centre d'Etudes Nucl\'eaires de Bordeaux Gradignan,
             UMR5797, F-33170 Gradignan, France}
\affiliation{CNRS, IN2P3,
             Centre d'Etudes Nucl\'eaires de Bordeaux Gradignan,
             UMR5797, F-33170 Gradignan, France}

\author{M. Bender}
\affiliation{Universit\'e Bordeaux,
             Centre d'Etudes Nucl\'eaires de Bordeaux Gradignan,
             UMR5797, F-33170 Gradignan, France}
\affiliation{CNRS, IN2P3,
             Centre d'Etudes Nucl\'eaires de Bordeaux Gradignan,
             UMR5797, F-33170 Gradignan, France}

\author{P.-H. Heenen}
\affiliation{PNTPM, Universit\'e Libre de Bruxelles, 1050 Bruxelles, Belgium}

\author{V. Hellemans}
\affiliation{PNTPM, Universit\'e Libre de Bruxelles, 1050 Bruxelles, Belgium}

\begin{abstract}

\begin{description}
\item[Background]
Symmetry restoration and configuration mixing in the spirit of the
generator coordinate method based on energy density functionals
have become widely used techniques in low-energy nuclear structure
physics. Recently, it has been pointed out that these techniques are
ill-defined for standard Skyrme functionals, and a regularization
procedure has been proposed to remove the resulting spuriosities
from such calculations.  This procedure imposes an integer power
of the density for the density dependent terms of the functional.
At present, only dated parameterizations of the Skyrme interaction
fulfill this condition.
\item[Purpose]
To construct a set of parameterizations of the Skyrme energy density
functional for multi-reference energy density functional calculations
with regularization using the state-of-the-art fitting protocols.
\item[Method]
The parameterizations were adjusted to reproduce ground state properties
of a selected set of doubly magic nuclei and properties of nuclear matter.
Subsequently, these parameter sets were validated against properties
of spherical and deformed nuclei.
\item[Results]
Our parameter sets successfully reproduce the experimental binding
energies and charge radii for a wide range of singly-magic nuclei.
Compared to the widely used SLy5 and to the SIII parameterization
that has integer powers of the density, a significant improvement
of the reproduction of the data is observed. Similarly, a good
description of the deformation properties at $A\sim 80$ was obtained.
\item[Conclusions]
We have constructed new Skyrme parameterizations with integer powers
of the density and validated them against a broad set of experimental
data for spherical and deformed nuclei. These parameterizations are
tailor-made for regularized multi-reference energy density functional
calculations and can be used to study correlations beyond the mean-field
in atomic nuclei.
\end{description}

\end{abstract}

\maketitle

%
%

\section{Introduction}

One of the most widely used designs of an effective nucleon-nucleon
interaction for mean-field-based methods~\cite{Ben03a}
was introduced by Skyrme~\cite{Sky56a,Sky58a} as a combination
of momentum-dependent two-body contact forces and a
momentum-independent three-body contact force. Already the first
applications~\cite{Vau72a,Vau73a,Bei75a} demonstrated the remarkable
qualities of this interaction to describe many properties of nuclei
throughout the chart of nuclei. However, some drawbacks
due to an insufficient flexibility of Skyrme's original \emph{ansatz}
became apparent.

The early parameterizations of Skyrme's interaction led to two major
problems. First, the simple contact three-body force does not allow
for a realistic value of the incompressibility $K_\infty$ of symmetric
infinite nuclear matter. Typically, values of about 350 MeV were
obtained, which is significantly larger than the empirical value of
$210 \lesssim K_\infty  \lesssim 240$~MeV
\cite{Bla76a,Bla80a,Col04a}.
Second, the same three-body force gives almost always rise to a
spin-instability in infinite nuclear matter \cite{Cha75a,Back75a,War76a}
and finite nuclei \cite{Str76a}, rendering the calculation of
excitations of unnatural parity in RPA impossible \cite{Bla76b}.

It turned out that both problems can be simultaneously solved when
replacing Skyrme's three-body force
$t_3 \, \delta(\vec{r}_1-\vec{r}_2) \delta(\vec{r}_1-\vec{r}_3)$
with a density-dependent two-body contact force
$\tfrac{1}{6}\, t_3 \, ( 1 + x_3 \hat{P}_\sigma) \,
\rho_0^\alpha[(\vec{r}_1+\vec{r}_2)/2] \,
\delta (\vec{r}_1-\vec{r}_2)$,
where $\hat{P}_\sigma$ and $\rho_0(\vec{r})$ are the spin-exchange operator
and the isoscalar density, respectively.
For $\alpha = 1$ and $x_3 = 1$, both are
equivalent as long as time-reversal symmetry is conserved~\cite{Vau72a}.
However, the so-called time-odd terms from the
density-dependent two-body force have a different isospin structure
than those of the three-body force, which removes the spin instability
\cite{Cha75a}. Therefore, all early parameterizations have since been
used as two-body forces with a linear density dependence.
In fact, the ambiguity around the three-body term was recognized
from the beginning by the authors of the earliest fits of Skyrme's interaction
\cite{Vau72a,Bei75a}, who pointed out that its three-body force
\emph{"should not be considered as a real three-body force, but
rather as convenient way of simulating the density dependence
of an effective interaction"} \cite{Bei75a}.

In a second
step, reducing the exponent $\alpha$ to values between $1/6$ and $1/3$
allows also for a realistic compressibility~\cite{Bei74a,Bla76a,Kri80a}.
The appearance of density dependencies of the form $\rho_0^\alpha (\vec{r})$
is also motivated through approximations to the $G$ matrix of
Brueckner-Goldstone theory~\cite{Koh69a,Bet71a,Koh75a,Koh76a}.
Up to now, all widely used parameterizations of the Skyrme interaction
have stuck to this simple form of density dependence, although several
extensions were attempted over time~\cite{Ben03a}. The same form of
density-dependent two-body contact force is also used to complement
the finite-range Gogny interaction~\cite{Gog75a,Dec80a}.

Modifications of specific terms in the total energy have been made as
well, hence abandoning the link to an underlying force \cite{Ben03a}.
Then, it is more appropriate to refer to a Skyrme energy density
functional (EDF).

There have been many adjustments of the parameters of Skyrme's interaction
since the 1970s~\cite{Ben03a}. The range of data on which the parameters
are fitted has been varied and extended, sometimes with choices dictated
by specific applications. Most fitting protocols, however, are designed
to deliver multi-purpose parameterizations that can be used for applications
as diverse as the description of ground-state masses and density
distributions, deformations, rotational bands, the response to external
probes, fission, and reaction dynamics for
nuclei all over the mass table and even the properties of neutron stars.
One of these is the protocol by Chabanat \emph{et al}.\ developed in the
1990s~\cite{Cha97a,Cha98a}, which led to a significant improvement of
isospin properties by including pseudo-data for neutron matter. The
resulting parameterizations, in particular SLy4, have been extensively
tested and used for the description of many properties of atomic nuclei
throughout the nuclear chart.

The Skyrme interaction was designed for use in self-consistent methods,
\emph{i.e.}\ Hartree-Fock (HF), HF+BCS, Hartree-Fock-Bogoliubov (HFB),
both in their static and time-dependent variants, and in RPA.
From the 1990s, Skyrme's interaction was also frequently employed in
extensions of mean-field methods, such as the construction of a
microscopic Bohr Hamiltonian \cite{Fle04a,Pro04a}, exact
projection \cite{Hee93a,Zdu07a} and configuration mixing by the
generator coordinate method (GCM)
\cite{Flo76a,Bon90a,Taj92a,Ben06a,Ben06b,Ben08a}.
Two issues were apparent from the beginning, one related to the adjustment
of the parameters of the interaction and the other to its analytical form.

Because the adjustment of the parameters is done within the mean-field
approximation, the inclusion of beyond-mean-field correlations
will often give rise to an overbinding
of nuclei, in particular of those used during the fit. The extra binding,
however, was always found to be within a few MeV and appears to
saturate quickly when several collective modes and symmetry restorations
are added consecutively \cite{Ben08a}. Here, we shall not consider this
issue as overbinding constitutes a small smooth trend that at the
present stage is smaller than other systematic errors and/or
uncertainties~\cite{Ben03a,Ben06a,Les07a,Kor08a,Ben09b}. Instead, we will
postpone the role of correlations on the outcome of a parameter
fit to future work and concentrate on the setup of the functional itself

As already mentioned, the Skyrme functional is adopted in extensions of the
mean-field approach. Such
 a functional
is \emph{a priori} defined only
for mean-field calculations, {\em i.e.} for a single mean-field
wave function, whereas beyond-mean-field calculations require
to determine a matrix element between wave functions generated by two
different mean fields. Unlike a formalism based on a
Hamiltonian, the extension of a density functional from a
single-reference (SR) definition to a multi-reference (MR) one is not an
unambiguous procedure. In the {early applications of
GCM using a Skyrme functional~\cite{Flo76a},
the SIII and SIV parameterizations were indeed used as two-
and three-body forces to calculate matrix elements between two
different mean-field wave functions.
Taking advantage of the generalized Wick theorem derived by Balian and
Br{\'e}zin~\cite{BB69a},
this amounts to replacing the mean-field densities that enter the energy
density by so-called \emph{mixed densities}. This scheme for the construction
of the energy density was also followed in Ref.~\cite{Bon90a},
where the parameterization SIII was adopted as a
density-dependent energy functional to construct the energy kernel in
a GCM calculation mixing mean-field states with different axial quadrupole
deformation. This procedure was not altered until very recently in subsequent
applications with more recent functionals of the Skyrme, Gogny and
relativistic type \cite{Ben06a,Ben08a,Rod02a,Yao10a} that often include
multiple symmetry restorations.

However, such a generalization of the functional ignores several
complications, in particular the fact that the {mixed densities can
become complex in MR calculations and that, often, different
functionals are chosen in the mean-field and  pairing channels.
Still,  the substitution} of mean-field
densities by mixed densities in the construction of the MR EDF
was used with some success in many applications, despite its drawbacks
that can in principle lead to unreliable results for an energy functional.
In one way or the other, the problems
of the functionals mentioned thus far are related to the breaking of the Pauli
principle~\cite{Ang01a,Dob07a,Lac09a}. In the standard Skyrme EDF, this has
many facets. First, the density dependence itself cannot be written
in a completely antisymmetrized form. Second, it is customary to use
different effective interactions for the particle-hole and
particle-particle parts of the EDF. In addition, certain exchange terms
of the Skyrme interaction are sometimes neglected or modified, and
for the Coulomb exchange term approximations are used. All of these
are either motivated by phenomenology, or by computational reasons.
For an overview, we refer to Ref.~\cite{Ben03a}. The standard density
dependence $\rho_0^\alpha(\vec{r})$ poses one additional problem. In
all currently used prescriptions, the density entering the
density-dependence might become complex. For non-integer
values of $\alpha$, the function $\rho_0^\alpha(\vec{r})$ then becomes a
non-analytical function of $\rho_0$ that is multivalued and exhibits
branch cuts~\cite{Ang01a,Dob07a,Dug09a}. To resolve this particular issue,
some alternatives for the density
dependence were formulated. Indeed, several studies have concentrated
on the most appropriate definition of the density dependence in MR
calculations, primarily for symmetry
restorations~\cite{Egi91a,Rod02a,Dug03a,Dob07a,Rob07a,Rob10a}.
The question, however, is not settled yet.

The net result of these problems is that the off-diagonal terms
in the MR EDF can exhibit discontinuities or even divergences when
varying one of the collective coordinates.
We refer to \cite{Ang01a,Dob07a,Lac09a,Ben09a,Dug09a} for an in-depth
analysis of these issues but present the arguments for the lack of
signs of their presence in the published GCM calculations.
First, the problems are especially critical for very light nuclei,
but applications were often devoted to medium-mass and heavy ones.
Second, the discretizations commonly chosen for numerical reasons
when setting up projection and GCM restrain the contamination of
the energy with non-physical contributions
to a very small scale.

One possibility to avoid these problems altogether would be a return to
a Skyrme-force-based Hamiltonian. This, however, will inevitably demand
the systematic addition of higher-order terms in the Skyrme force, as
within the standard form it is impossible to construct a parameterization
that, at the same time, describes the empirical properties of nuclear
matter, has no
spin or other instabilities, and gives attractive pairing. By
contrast, within an energy functional framework a fair description of
nuclear matter and finite nuclei is achieved within the standard form.
Thus, to keep the effective interaction simple, it appears to be
preferable to work with a functional instead of a force. To enable
their use in a MR framework, tools to by-pass the obstacles
outlined above by a regularization of the Skyrme functional have been
designed recently~\cite{Lac09a,Ben09a}. They require, however, that the
functional dependence on the density has
an integer power~\cite{Dug09a}.

In this article, we construct Skyrme functionals that have the same
density dependence as SIII and thereby are regularizable in the sense of
Ref.~\cite{Lac09a}. The first parameterizations of the Skyrme functional
built about 40 years ago~\cite{Bei75a} had all this property, but, since
then, the fitting protocols
have significantly evolved and these early parameterizations certainly have
to be reconsidered. Our study is based on the protocol first used for
the SLy$x$ parameterizations~\cite{Cha97a,Cha98a} that has proven to
be efficient to construct functionals used successfully in a large number
of applications. In this first study, we will restrict ourselves to
the standard form of the Skyrme functional. The construction of a
regularizable functional including higher-order density-dependent terms
is underway \cite{Sad11t,Sad12x} and will be reported elsewhere.
However, we take the opportunity of the present study to include a new
set of data in the fitting protocol, which are used to
validate (or reject) the parameterizations.

There is a major conceptual difference between the
parameterization of the Skyrme functional that we aim at and the ones by
Kortelainen \emph{et al}.~\cite{Kor10a,Kor11a}, who have recently
adjusted new Skyrme parameterizations on a large set of data. The aim of
Kortelainen \emph{et al}.\ is to describe the nucleus in the spirit of
the density functional theory \cite{Eng11a} that is very successful
in condensed matter physics. Staying on the computationally simple
single-reference level, as much correlation energy as possible is
incorporated into the energy functional. Our aim is to construct
a parameterization of the Skyrme EDF that will be used in
beyond-mean-field calculations, \emph{i.e.}\ where specific correlations
are to be calculated explicitely in a multi-reference framework.
Both views are
complementary. The advantage of our approach is that it enables
to calculate spectra and transition probabilities directly in the
laboratory frame of reference and avoids the ambiguities related
to approximate determinations of spectroscopic quantities, whereas
its disadvantage is that already for standard observables
high predictive power will require the time-consuming calculation of
correlations beyond the mean field. In the following, we will call
beyond mean-field method the method that we have already used in
many applications and where mean-field wave functions generated by
a constraint on a collective variable are projected on particle numbers
and angular momentum and mixed by the GCM.

The article is organized as follows. Section~\ref{sec:protocol}
reviews the fitting protocol used here and its differences to the
one used to construct the SLy$x$ parameterizations in the past. In
Sec.~\ref{sec:results}, we will test the parameterizations on a
large set of typical observables for spherical and deformed nuclei,
including masses, separation energies, charge radii, deformations,
the fission barrier of \nuc{240}{Pu}, and the moment of inertia of
a superdeformed rotational band in \nuc{194}{Hg}. Section~\ref{sec:summary}
will summarize our findings.

%
%
\section{Fitting protocol}
\label{sec:protocol}
%
%

\subsection{The energy functional}
\label{subsec:EDF}

The standard density-dependent Skyrme interaction has the form~\cite{Les07a}
\begin{eqnarray}
\label{eq:vskyrme}
v (\vec{R},\vec{r})
& = &  t_0 \, ( 1 + x_0 \hat{P}_\sigma ) \; \delta (\vec{r})
      \nn \\
&   & + \tfrac{1}{6} \, t_{3} \, ( 1 + x_{3} \hat{P}_\sigma ) \,
        \rho^{\alpha}_0 (\vec{R}) \; \delta (\vec{r})
      \nn \\
&   & + \tfrac{1}{2} \, t_1 \, ( 1 + x_1 \hat{P}_\sigma )
        \big[   \hat{\vec{k}}^{\prime 2} \; \delta (\vec{r})
              + \delta (\vec{r}) \; \hat{\vec{k}}^2
        \big]
      \nn \\
&   & + t_2 \ ( 1 + x_2 \hat{P}_\sigma ) \,
      \hat{\vec{k}}^{\prime} \cdot \delta (\vec{r}) \; \hat{\vec{k}}
      \nn \\
&   & + \mathrm{i}\, W_0 \, ( \hat{\vsigma}_1 + \hat{\vsigma}_2 ) \cdot
      \hat{\vec{k}}^{\prime} \times \delta (\vec{r}) \; \hat{\vec{k}}
\, ,
\end{eqnarray}
where we use the shorthand notation $\vec{r} \equiv \vec{r}_1 - \vec{r}_2$
and $\vec{R} \equiv \tfrac{1}{2} ( \vec{r}_1 + \vec{r}_2 )$ for the relative
distance and center-of-mass coordinates, respectively, where
$\hat{P}_\sigma$ is the spin exchange operator,
$\hat{\vec{k}} \equiv -\frac{\mathrm{i}}{2}(\vnabla_1-\vnabla_2)$ the
relative momentum operator acting to the right,
and $\hat{\vec{k}}'$ is the complex conjugate
of $\hat{\vec{k}}$ acting to the left, and $\rho_0 (\boldsymbol{R})$ is
the isoscalar density. The Skyrme interaction~(\ref{eq:vskyrme}) contains
in total 10 parameters $t_0$, $t_1$, $t_2$, $t_3$, $x_0$, $x_1$, $x_2$,
$x_3$, $W_0$, and $\alpha$ to be adjusted to data.

As it is customary, we only calculate the particle-hole part of the EDF
from Eq.~(\ref{eq:vskyrme}). We keep, however, all terms in that channel,
which is not always done~\cite{Ben03a}. For the special case of
time-reversal invariance and spherical symmetry this leads to
\begin{multline}
\label{eq:EDF}
{\cal E}_{\rm Skyrme}
= \int \! d^3r \sum_{t=0,1}
  \Bigl\{ C_t^{\rho}[\rho_0] \rho_t^2 +
          + C_t^{\Delta\rho}\rho_t\Delta\rho_t  \\
          + C_t^\tau\rho_t\tau_t
          +\frac{1}{2} C_t^J\boldsymbol{J}_t^2
          + C_t^{\nabla\cdot J}\rho_t\nabla\cdot\boldsymbol{J}_t
  \Bigr\} \, ,
\end{multline}
where $\rho$, $\tau$, and $\boldsymbol{J}$ are the density, kinetic
density, and spin-current vector density, respectively, and the index
$t$ labels isoscalar ($t=0)$ and isovector ($t=1)$ densities.
The definition of these densities and the relations between the
coefficients $C_t$ in Eq.~(\ref{eq:EDF}) and the parameters
in Eq.~(\ref{eq:vskyrme}) can be found in Ref.~\cite{Les07a}.
Note that the coefficients $C_t^{\rho}[\rho_0]$ depend on the
isoscalar density $\rho_0 (\boldsymbol{r})$, whereas all others
are just numbers. In case of deformed nuclei and when breaking intrinsic
time-reversal symmetry, there are additional terms in the Skyrme EDF
for which we refer to Refs.~\cite{Les07a,Hel12a}

The total energy is given by the sum of the Skyrme EDF~(\ref{eq:EDF}),
the Coulomb energy, the kinetic energy, the center-of-mass correction
and the pairing energy. As in our previous studies, we have chosen a
density-dependent zero-range pairing interaction~\cite{Ter95a,Ben03a,Hel12a},
which leads to a functional of the form
\begin{equation}
\label{eq:Epair}
\mathcal{E}_{\text{pairing}}
= \frac{V_0}{4} \sum_{q=p,n}
  \int \! d^{3}\vec{r}
  \left[1-\frac{\rho_{0}(\vec{r})}{\rho_{c}}\right]
  \tilde{\rho}_{q}(\vec{r}) \, \tilde{\rho}_{q}^{\ast}(\vec{r})
\, .
\end{equation}
The switching density $\rho_{c} = 0.16$ fm$^{-3}$ is set to the empirical
nuclear saturation density, such that the pairing interaction is most
active on the surface of the nucleus. The pairing functional depends on
the local pair density $\tilde{\rho}_{q}(\vec{r})$~\cite{Hel12a} of
protons and neutrons, labeled by $q = p$, $n$, and the isoscalar local density
$\rho_{0}(\vec{r})$. An energy cutoff of 5~MeV in the single-particle
spectrum is taken above and below the Fermi energy~\cite{Kri90a}.
The strength $V_0$ will be adjusted separately for each parameterization
of the Skyrme interaction.

For most (if not all) Skyrme interactions constructed up to now, the
Coulomb exchange energy has been replaced by its Slater approximation
\cite{Ben03a} that amounts to a local energy density of the form
$\sim \rho_p^{4/3}(\boldsymbol{r})$,  {\em i.e.} a term depending on a
non-integer power of the density. Like the standard density dependence
in the Skyrme EDF $\sim \rho_0^\alpha(\boldsymbol{r})$ with
$0 < \alpha < 1$, this term cannot be regularized with the currently
available techniques~\cite{Dug09a}. For interactions that can be safely
used in regularized MR~EDF calculations, the Coulomb exchange energy
has to be either treated exactly or to be omitted. For simplicity,
we have chosen to neglect it in the mean-field channel in the present study since an exact
treatment of the Coulomb exchange field
makes all calculations much
more time consuming. In addition, phenomenological arguments have also
been brought forward that justify this course of
action~\cite{Bro98a,Bro00a,Gor08a}.
As usually done, the contribution of the Coulomb interaction to the pairing channel is neglected.

For the center-of-mass correction, we employ the widely-used approximation
where only the one-body term is considered~\cite{Ben00b}. However,
the often neglected $\boldsymbol{J}^2$ term in the Skyrme
functional~(\ref{eq:EDF}) is kept. The latter two choices correspond to
the ones made for the parameterization SLy5 of Chabanat
\emph{et al.}~\cite{Cha98a}.

%
%
\subsection{The protocol}
\label{subsec:protocol}

The  first step of our fitting protocol is similar to the one used
for the construction of the SLy$x$ parameterizations~\cite{Cha97a,Cha98a}.
During this step, we minimize a merit function which is a weighted sum
of squared residuals:
\begin{equation}
\label{eq:chisquare}
\chi^2
= \sum_A \chi^2_A \, , \quad
\chi^2_A
= \frac{1}{N_A}\sum_{i=1}^{N_A}
  \left( \frac{O_i-O_i^{\rm calc.}}{\Delta O_i} \right)^2
\, .
\end{equation}
The $O_i$ are experimental data for finite nuclei and empirical values for
nuclear matter and the $\Delta O_i$ are tolerance parameters used to weight
these data during the fit. Five categories of data are used:
\begin{enumerate}
  \item nuclear matter properties around the saturation point,
  \item neutron matter equation of state,
  \item binding energies of doubly-magic nuclei,
  \item charge radii,
  \item spin-orbit splittings of neutron and proton states.
\end{enumerate}
The nuclear matter properties that we have included are:
\begin{itemize}
\item
the saturation density $\rho_\mathrm{sat} = 0.16$~fm$^{-3}$
with a tolerance $\Delta O_i= 0.003~\mathrm{fm}^{-3}$;
\item
the binding energy per nucleon $E/A = -16$~MeV with
$\Delta O_i = 0.3~\mathrm{MeV}$;
\item
the symmetry energy $a_{\rm sym}=31$~MeV with $\Delta O_i=1~\mathrm{MeV}$;
\item the Thomas--Reiche--Kuhn sum rule enhancement factor $\kappa_v=0.25$
with $\Delta O_i= 0.15$.
\end{itemize}
Since the incompressibility of nuclear matter $K_\infty$ cannot be
adjusted to a realistic value with the restriction imposed on
$\alpha = 1$ \cite{Cha97a}, this quantity is not considered in
our fitting protocol.

The binding energies of six doubly-magic nuclei are included:
$^{48}$Ca, $^{132}$Sn, and $^{208}$Pb with tolerances of
$\Delta O_i = 0.2$~MeV, $^{40}$Ca and $^{100}$Sn with
$\Delta O_i = 0.5$~MeV, and $^{56}$Ni with $\Delta O_i=0.75$~MeV.
We allow for larger $\Delta O_i$ for $N=Z$ nuclei as one always
has difficulties to reproduce their binding energy at the
mean-field level. However, the discrepancies cannot be simply
related to the Wigner energy that cannot be described by mean-field
calculations. Usually only $^{56}$Ni turns out to be underbound,
whereas $^{40}$Ca and $^{100}$Sn are overbound.
The charge radii of $^{40,48}$Ca, $^{56}$Ni, $^{132}$Sn,
$^{208}$Pb have a tolerance $\Delta O_i=0.02$~fm, and the
spin-orbit splittings of the neutron $3p$ levels and the proton
$1h$ levels in $^{208}$Pb have both a tolerance $\Delta O_i=0.2$~MeV.

For the neutron matter equation of state, $O_i$ are the
energies per neutron for $\rho \leqslant 0.5$~fm$^{-3}$
predicted by Wiringa \emph{et al}.~\cite{Wir88a} with the
bare two-body UV14 potential and three-body UVII potential.
The tolerance parameters are set to $\Delta O_i = 0.2\times O_i$.

These data are used to determine a first set of values of the
Skyrme parameterization. The resulting EDF is then tested on
several properties of finite nuclei that will be discussed
in the following sections. Among these properties, the charge
radii were strongly underestimated with the first set of
weights that we have used. We have therefore chosen to relax
the weights of nuclear matter properties, especially, the
density at saturation and the constraints on neutron matter
properties. After some
attempts, this was sufficient to
arrive to a satisfactory reproduction of charge radii.
The weights that are given above are the final weights used
in the fit.

During the first attempts to fit our new parameterizations,
we encountered finite-size isospin instabilities that are characterized
by a separation of protons and neutrons as examined in Ref.~\cite{Les06a}.
The instability appears when the coupling constant
$C_{1}^{\Delta \rho} = \frac{3}{32}t_1\left(\frac{1}{2}+x_1\right)
+\frac{1}{32}t_2\left(\frac{1}{2}+x_2\right)$ in the Skyrme
EDF~(\ref{eq:EDF}) takes too large a value. To prevent such
instabilities, we enforce a condition on the coupling constant
\begin{eqnarray}
\label{eq:constrc1}
\chi^2_{A}
=\left\{\begin{array}{cc}
 \left(\frac{  C_{\rm 1,calc}^{\Delta \rho}
             - C_{\rm 1,max}^{\Delta \rho}}{1.5} \right)^2
        &\text{for $C_{\rm 1,calc}^{\Delta \rho} \geqslant C_{\rm 1,max}^{\Delta \rho}$}, \\
0 &\text{for $C_{\rm 1,calc}^{\Delta \rho} < C_{\rm 1,max}^{\Delta \rho}$},
\end{array}\right.
\end{eqnarray}
where the empirical choice for the maximum value
$C_{\rm 1,max}^{\Delta \rho} = 25$~MeV\,fm$^5$ has been found to lie safely
within the stable zone. We have also checked that the parameterizations
do not lead to finite-size instabilities due to the
$\vec{s}_t \cdot \Delta \vec{s}_t$ terms in the time-odd part of the
Skyrme EDF~\cite{Hel12a} when setting the corresponding coupling constants
to their Skyrme force value.

%
\section{Results}
\label{sec:results}

%
%
\subsection{New parameter sets}
\label{subsec:parameters}

\begin{table}[t!]
\caption{
\label{table:parameters}
New parameter sets for the Skyrme energy functional with effective masses
as indicated.
}
\begin{ruledtabular}\begin{tabular}{lrrrr}
\noalign{\smallskip}
 &  \multicolumn{1}{c}{0.7}
 &  \multicolumn{1}{c}{0.8}
 &  \multicolumn{1}{c}{0.9}
 &  \multicolumn{1}{c}{1.0}
   \\
\noalign{\smallskip}  \hline \noalign{\smallskip}
$t_0$ (MeV\,fm$^3$)&$-1122.408$&$-1100.272$&$-1082.609$&$-1066.976$ \\
$t_1$ (MeV\,fm$^5$)&  440.572  &  359.568  &   295.999 & 245.431    \\
$t_2$ (MeV\,fm$^5$)& $-197.528$& $-210.840$& $-240.653$&$-245.314$ \\
$t_3$ (MeV\,fm$^{6}$)&11906.299& 13653.845& 15003.161& 16026.086 \\
\noalign{\smallskip}  \hline \noalign{\smallskip}
$x_0$ & 0.394119  & 0.445280  & 0.491775  & 0.525497        \\
$x_1$ &0.068384  & 0.224693  & 0.389884  & 0.603399     \\
$x_2$ &$-0.752728$&$-0.615015$&$-0.579284$&$-0.500115$     \\
$x_3$ & 0.946945  & 0.639947  & 0.512106  & 0.366056        \\
\noalign{\smallskip} \hline \noalign{\smallskip}
$W_0$(MeV\,fm$^5$)& 119.125   & 110.828   & 103.516   & 97.977  \\
$\alpha$          & 1         & 1         & 1         & 1
\end{tabular}\end{ruledtabular}\end{table}

\begin{table}[t!]
\caption{
\label{table:nuclearmatter}
Saturation properties of nuclear matter as obtained with the
new parameter sets. Values for SIII and SLy5 are shown for comparison.
}
\begin{ruledtabular}
\begin{tabular}{lcccccc}
 &$ 0.7 $&$0.8 $&$ 0.9 $&$ 1.0 $ & SIII & SLy5   \\
\noalign{\smallskip} \hline \noalign{\smallskip}
$\rho_\mathrm{sat}$ (fm$^{-3}$)
  & 0.153       & 0.153 & 0.153  & 0.153 & 0.145  & 0.160\\
$E/A$ (MeV)     & -16.33 & -16.32 & -16.31 & -16.31 & -15.85 & -15.98 \\
$m^*_0/m$       & 0.700     & 0.800  & 0.900  & 1.000  & 0.763 & 0.697 \\
$K_\infty$ (MeV)    & 361.3  & 368.7 & 374.5   & 379.4    & 355.4 & 229.9 \\
$a_{\rm sym}$ (MeV) & 31.98 & 31.69  & 31.44  & 31.31  & 28.16 & 32.03 \\
$\kappa_v$          & 0.612 & 0.467  & 0.336   & 0.250   & 0.525 & 0.250
\end{tabular}
\end{ruledtabular}
\end{table}

\begin{figure}[t!]
\includegraphics[width=0.9\linewidth, clip]{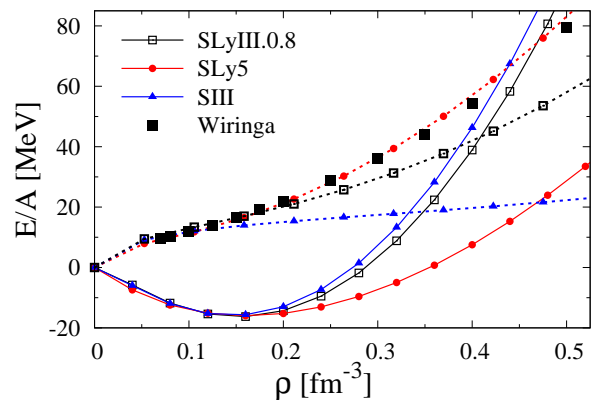}
\caption{
\label{fig:eos}
(Color online)
Binding energy per nucleon $E/A$ for symmetric nuclear matter (solid lines)
and for pure neutron matter (dotted lines) as a function of nucleon
density $\rho$. The filled squares indicate empirical values of the neutron
matter equation of state by Wiringa \emph{et al.}~\cite{Wir88a}.
}
\end{figure}

The fact that we do not constrain the compressibility of nuclear
matter leaves some freedom in the choice of the effective mass,
cf.\ the discussion in Ref.~\cite{Cha97a}.
We have constructed four parameter sets corresponding to
values of the isoscalar effective mass $m^*_0$ from 0.7 to $1.0$
times the nucleon mass $m$. We will refer to these as SLyIII.$xx$,
where $xx$ is the value of $m^*_0/m$.

The coupling constants of these four parameterizations are listed
in Table~\ref{table:parameters}, and the corresponding saturation
properties of infinite homogeneous nuclear matter in
Table~\ref{table:nuclearmatter}.
As expected, the value of $K_\infty$ is much too large. It increases
with the effective mass~\cite{Cha97a} and there is no room to obtain
a value close to the empirical value when imposing $\alpha=1$ without introducing
additional terms in the Skyrme functional. To obtain a
reasonable
agreement between theory and experiment for charge radii has required
to relax the constraint on $\rho_\mathrm{sat}$ leading to a value lower
than the usual one of $0.160$~fm$^{-3}$, but still larger
than the one for SIII.

The equation of state $E/A$ of symmetric infinite matter obtained with
SLyIII.0.8 is compared with results for SLy5 and SIII in Fig.~\ref{fig:eos}.
As can be expected from the values for $K_\infty$, it is stiffer than
the equation of state obtained with SLy5.

In the same figure, we also compare the binding energy per
neutron for pure neutron matter determined using SLy5, SIII and SLyIII.0.8
to \emph{ab-initio} results obtained by Wiringa
\emph{et al.}~\cite{Wir88a}.
On the scale of the plot, obvious differences between the
parameterizations appear only at rather large densities
$\rho_n \gtrsim 0.12$~fm$^{-3}$. At values below,
the results obtained with the three parameterizations cannot be
easily distinguished, in spite of the fact that SIII was not fitted
to this quantity. For larger densities, however, as expected, the
inclusion of the neutron matter equation of state in the fitting
protocol improves the results obtained with SLyIII.xx with respect
to those of SIII. For SLy5, the tolerance in the merit function,
Eq.~(\ref{eq:chisquare}) has been chosen much smaller
than for the SLyIII.$xx$, leading to a better reproduction of the equation
of state.

The residuals of binding energies and charge radii of doubly-magic
nuclei are displayed in Fig.~\ref{fig:doublemagic} and the
corresponding values of $\chi^2$ are given in Table~\ref{table:chi2}.
In both cases, the new parameterizations perform much better than SIII
and SLy5, irrespective of the value of the effective mass.
We have to recall, however, that SIII and SLy5
were fitted with different protocols, such that the comparison of the
$\chi^2$ can only serve as a guideline for the relative performance of
the parameterizations for these specific observables. It does not allow
to judge their overall quality. In particular, as discussed above, SLy5
gives a much better description of some key nuclear matter properties
that cannot be adjusted with SLyIII.$xx$.

\begin{table}[t!]
\caption{
\label{table:chi2}
$\chi^2$ values in Eq.~(\ref{eq:chisquare}) for binding
energies $E$ and charge radii $r_c$.
}
\begin{ruledtabular}
\begin{tabular}{crrrrrr}
      & 0.7 & 0.8 & 0.9  & 1.0  & SIII & SLy5 \\
\noalign{\smallskip} \hline \noalign{\smallskip}
$E$   &  5.12 & 4.33 & 3.93 & 4.02 & 63.99 & 14.80  \\

$r_c$ & 0.67  & 0.74 & 0.87 & 1.22 & 7.79  & 1.74
\end{tabular}
\end{ruledtabular}
\end{table}

\begin{figure}[t!]
\includegraphics[width=\linewidth, clip]{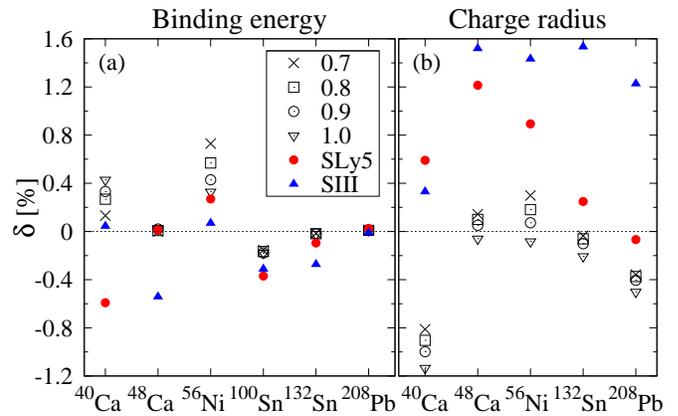}
\caption{(Color online)
Residuals between calculated and experimental
values~\cite{Aud03a,Nad94a,Bla05a},
defined as $\delta=(O^{\rm calc.}-O^{\rm expt.})/|O^{\rm expt.}| $,
for binding energies and charge radii of doubly-magic nuclei.
}\label{fig:doublemagic}
\end{figure}

%
%
\subsubsection{Adjustment of the pairing strength}

\begin{table}[t!]
\caption{
\label{table:strength}
Pairing strength $V_0$ for the parameterizations as indicated. The
switching density is set to $0.16$~fm$^{-3}$ in all cases.
}
\begin{ruledtabular}
\begin{tabular}{ccccccc}
     & 0.7 & 0.8 & 0.9 & 1.0  & SLy5 & SIII \\
\noalign{\smallskip} \hline \noalign{\smallskip}
$V_0$ (MeV\,fm$^3$) & 994& 987& 985& 988& 977 & 944
\end{tabular}
\end{ruledtabular}
\end{table}

To compute spherical and deformed open-shell nuclei, pairing correlations
need to be taken into account. The functional form and the adjustment
of a pairing interaction is a problem that requires, in principle,
a dedicated study of its own~\cite{Ben00a,Yam09a}. Since our focus is
on the properties of the interaction used in the particle-hole channel, we
restrict ourselves to the surface pairing
energy density functional~(\ref{eq:EDF}) that we have used in numerous
past studies.

The pairing strength $V_0$ in Eq.~(\ref{eq:EDF}) is fitted in \nuc{120}{Sn}
on the neutron spectral pairing gap
\mbox{$ -\mathcal{E}_{\text{pairing},n}
/ \int \! d^3r \; \tilde{\rho}_n (\vec{r})$} \cite{Ben00a,Yam09a}.
In this expression, $\mathcal{E}_{\text{pairing},n}$ is the pairing energy
of the neutrons and $\tilde{\rho}_n (\vec{r})$ the neutron pair density,
respectively.
The empirical value is determined by a five-point formula
for the gap~\cite{Ben00a} and is equal to 1.393~MeV. The pairing strengths
obtained for the four values of the effective mass are listed in
Tab.~\ref{table:strength}. They are very close to each other and do
not scale significantly with the effective mass.

%
%

\subsection{Spherical nuclei}
\label{subsec:spherical}
We start our validation of the SLyIII.xx interactions by confronting
their predictions with various experimental data for
singly-magic nuclei.

%
%
\subsubsection{Binding energies}

\begin{table}[t!]
\caption{
\label{table:err_E_r2}
Mean deviation with respect to the data for binding energies
and charge radii obtained for the parameterizations as indicated.
}
\begin{ruledtabular}
\begin{tabular}{lcccc}
\noalign{\smallskip}
  & $E_\mathrm{dev}$~(MeV)
  & $E_\mathrm{rel}$~(\%)
  & $r^c_\mathrm{dev}$~(fm)
  & $r^c_\mathrm{rel}$~(\%) \\
\noalign{\smallskip} \hline \noalign{\smallskip}
SLyIII.0.7 & 1.97 & 0.26 &  0.018 &   0.40  \\
SLyIII.0.8 & 1.46 & 0.21 &  0.020 &   0.46  \\
SLyIII.0.9 & 1.09 & 0.17 &  0.023 &   0.52  \\
SLyIII.1.0 & 0.98 & 0.15 &  0.029 &   0.65  \\
SLy5       & 2.49 & 0.31 &  0.012 &   0.29  \\
SIII       & 1.88 & 0.23 &  0.051 &   1.09
\end{tabular}
\end{ruledtabular}
\end{table}

\begin{figure*}[t!]
\includegraphics[width=0.75\linewidth, clip]{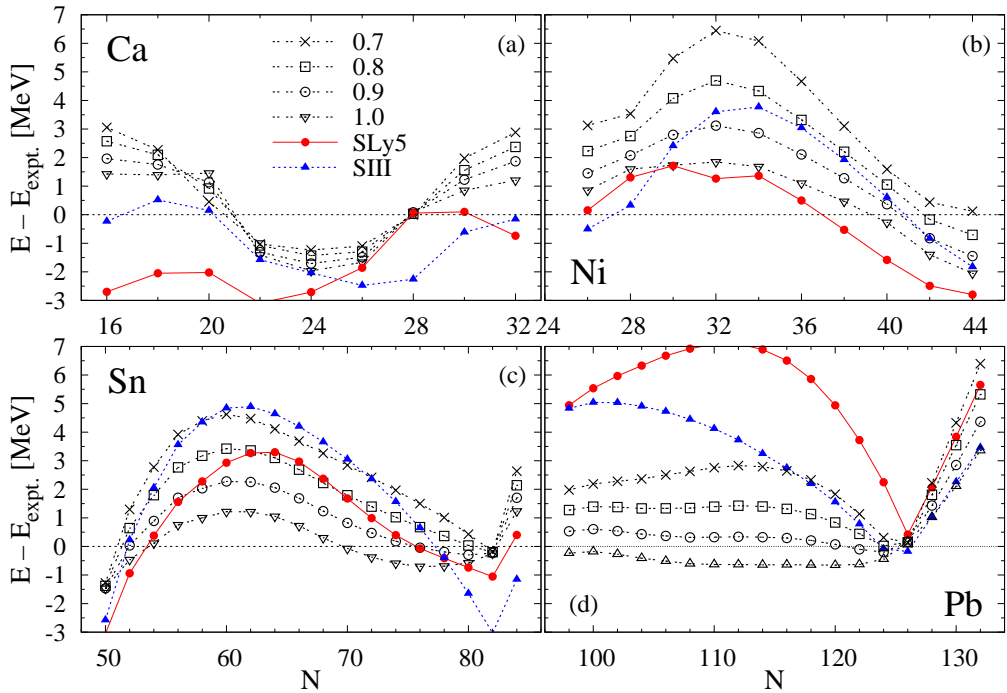}
\caption{
(Color online)
Residuals of the binding energy $E-E_{\rm expt.}$ as a function
of neutron number $N$ for the Ca, Sn, Ni, and Pb isotopic chains
obtained with the parameterizations as indicated.
Experimental data are taken from Ref.~\cite{Aud03a}.
}
\label{fig:energy_isotope}
\end{figure*}

\begin{figure*}[t!]
\includegraphics[width=0.75\linewidth, clip]{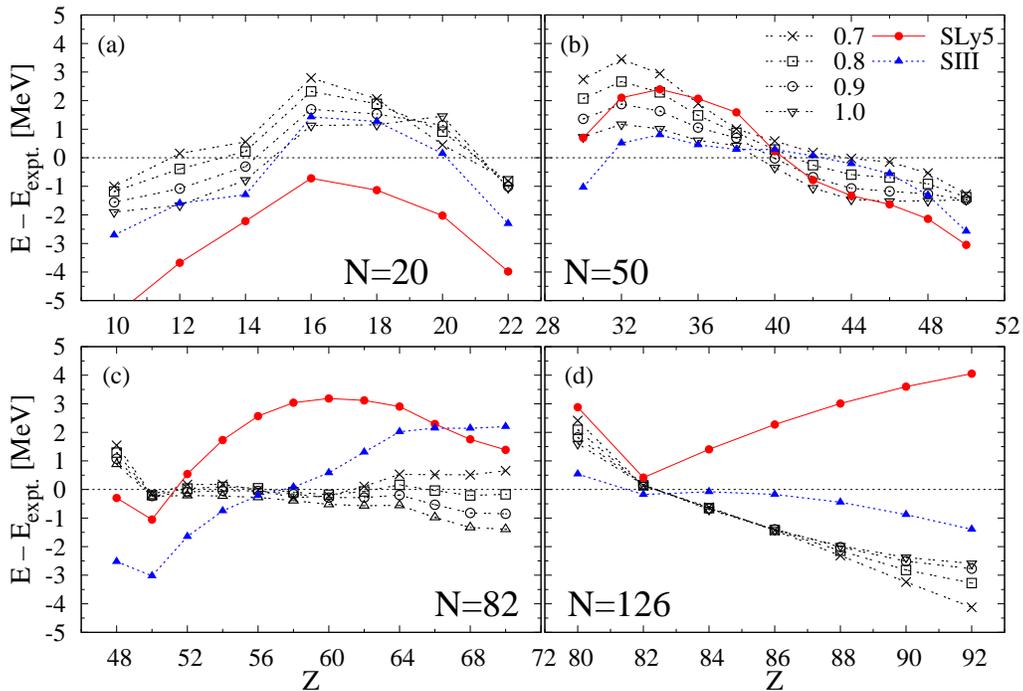}
\caption{
\label{fig:energy_isotone}
(Color online)
Same as in Fig.~\ref{fig:energy_isotope} but for $N=20$, 50, 82, 126
isotonic chains as a function of proton number $Z$.
}
\end{figure*}

The differences between the calculated and the experimental binding
energies are shown in Figs.~\ref{fig:energy_isotope}
and~\ref{fig:energy_isotone} for representative isotopic and
isotonic chains of singly-magic nuclei. The agreement with the data is
in general better for the SLyIII.$xx$ parameterizations than for
SIII and SLy5. To quantify these energy differences, we have defined
two mean deviations:
\begin{eqnarray}
E_\mathrm{dev}
& = & \frac{1}{N}
      \sum_{i=1}^N \big| E_i - E_i^\mathrm{expt.} \big| \, ,  \label{eq:Edev}\\
E_\mathrm{rel}
& = & \frac{1}{N}
      \sum_{i=1}^N \frac{ |E_i - E_i^\mathrm{expt.}|}{|E_i^\mathrm{expt.}| }
       \, ,
      \label{eq:Erel}
\end{eqnarray}
where $N$ is the total number of singly-magic nuclei that have
been calculated. Analogous quantities can be defined for charge radii.
The values given in Table~\ref{table:err_E_r2}
confirm that the agreement with data is improved
by the SLyIII.$xx$ parameterizations.  Deviations for binding energies
$E_{\mathrm{dev}}$ decrease with increasing effective mass.

Let us recall that our aim is to construct an interaction well suited
for adding the correlations generated by symmetry restorations and
configuration mixing calculations. Therefore, the nuclei calculated
at the mean-field level of approximation should be underbound, and
that in such a manner that the difference between mean-field calculation
and data is slightly larger for mid-shell nuclei than for doubly-magic
ones~\cite{Ben06a}.

It is clear that the SLyIII.\textit{xx} parameterizations with the largest
values of $m_0^*/m$ leave nearly no room for the addition of correlation
energies in the Sn and Pb chains. The increase of the effective mass
washes out the shell effects in the mean-field results. At this point,
SLyIII $m_{0}^{*}/m=0.8$ is the most promising parameterization,
underbinding the energy of the Sn and Pb isotopes by what can be expected
to be added from correlations.

%

\subsubsection{Charge radii}
\begin{figure}[t!]
\includegraphics[width=\linewidth, clip]{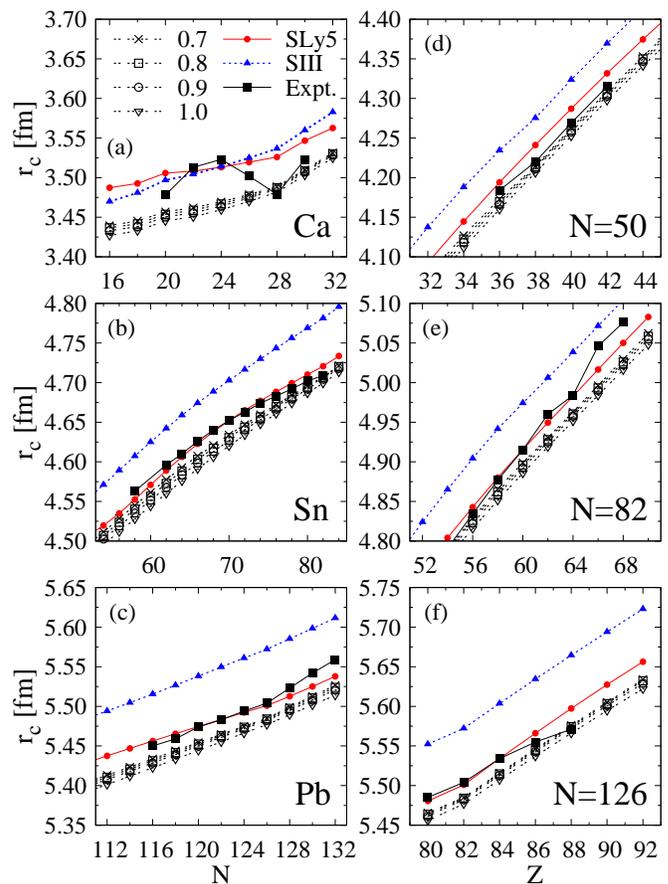}
\caption{\label{fig:rms}
(Color online)
Charge radii for Ca, Sn, Pb isotopic chains and for $N=50$, 82, 126
isotonic chains obtained by SLyIII, SLy5, and SIII. Experimental
data shown by solid squares are taken from~\cite{dataset04}.
}
\end{figure}

The calculated and experimental charge radii are compared in
Fig.~\ref{fig:rms}.  The charge radii are determined according to
Ref.~\cite{Cha97a}, taking into account the internal charge distribution
of both protons and neutrons and adding a correction for the
electromagnetic spin-orbit effect. The corresponding  deviations,
defined in Eqs.~\eqref{eq:Edev} and~\eqref{eq:Erel}, are given
in Table~\ref{table:err_E_r2}. The SLyIII.\textit{xx} parameterizations
clearly provide a better description of these data than SIII,
which systematically underestimates the charge radii. SLy5, on the other
hand, leads to even larger radii and therefore performs in general
better than the SLyIII.\textit{xx}. As can be seen from
Table~\ref{table:err_E_r2}, the deviations from the data $R_{\mathrm{dev}}$
and $R_{\mathrm{rel}}$ decrease with decreasing effective mass.

Again, we recall that correlations from fluctuations in
the quadrupole degree of freedom consistently increase the charge radii
of spherical nuclei~\cite{Ben06a}. Overall, the observed trend of the
charge radii is well reproduced by the calculation. The deviations
from the smooth trend observed in the data
for the Pb and Ca isotopes and for the $N=82$, and 126 isotones
are not described by any of the parameterizations and seemingly
require either the inclusion of explicit correlations, or higher-order
terms in the EDF, cf.\ Ref.~\cite{Ben03a} and references therein.

%
%

\subsubsection{Two-neutron separation energies}

\begin{figure}[t!]
\includegraphics[width=\linewidth, clip]{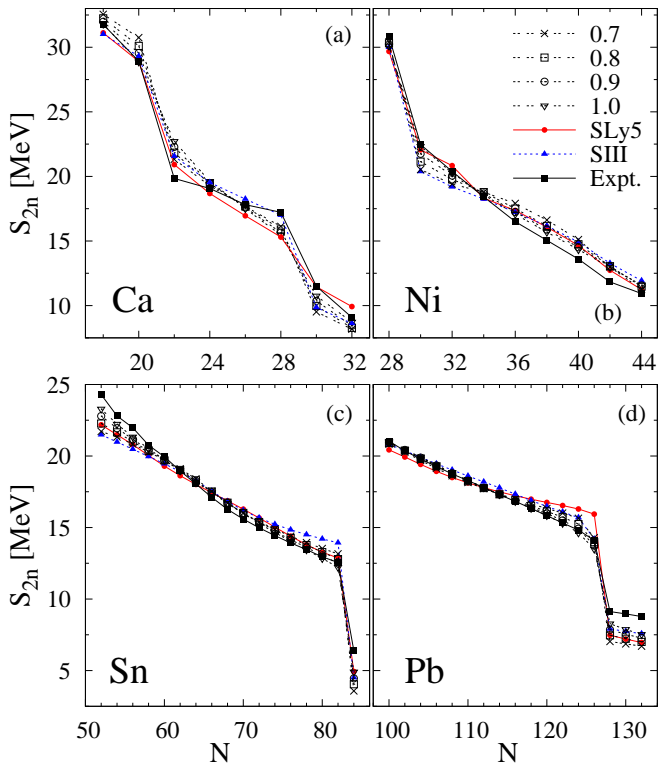}
\caption{
\label{fig:S2n_isotope}
(Color online)
Two-neutron separation energies for Ca, Ni, Sn, and Pb isotopes
calculated with the parameterizations as indicated.
}
\end{figure}

The two-neutron separation energies are compared to the experimental
data in Fig.~\ref{fig:S2n_isotope} for the Ca, Ni, Sn, and Pb
isotopic chains. All six parameterizations give similar results
for mid-shell nuclei. They tend to overestimate the characteristic
jump at neutron magic numbers, which, however, would be reduced by
dynamical quadrupole correlations~\cite{Ben06a}. For Ca
and Ni isotopes, our values do not reproduce the slope
of the experimental data for mid-shell nuclei. For the Sn and Pb
isotopes, the agreement with the data is improved by
SLyIII.\textit{xx} with respect to SIII and SLy5.

%
%

\subsubsection{Single-particle energies}

\begin{figure*}[t!]
\centerline{
\includegraphics[width=\linewidth, clip]{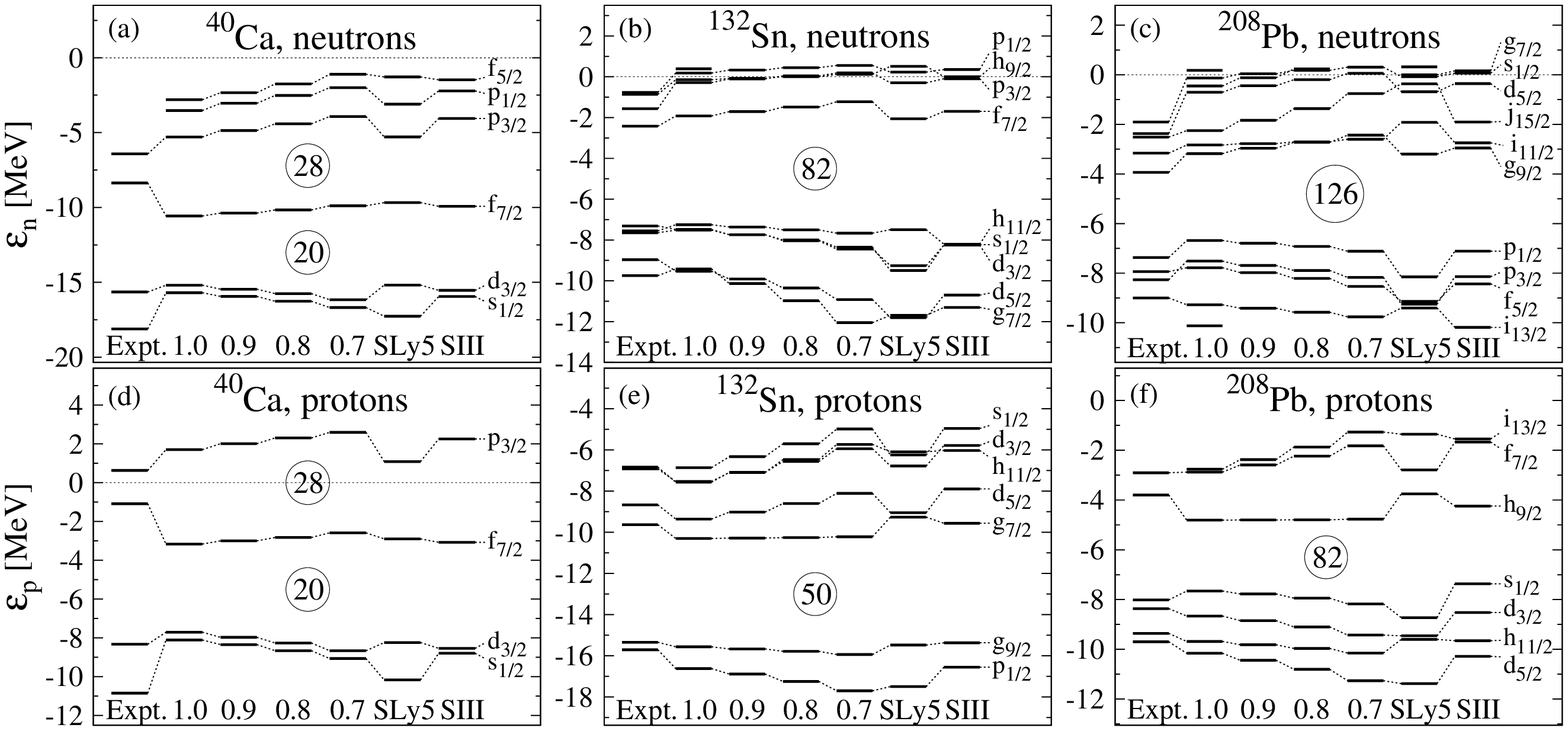}
}
\caption{
\label{fig:spenergy}
Single-particle energies obtained as eigenvalues of the mean-field
Hamiltonian in spherical calculations
for $^{40}$Ca (left), $^{132}$Sn (center), and $^{208}$Pb (right)
for the parameterizations as indicated and compared
with experimental data determined as one-nucleon separation
energies to or from the doubly-magic nucleus.
}
\end{figure*}

Up to now, our analysis of the Skyrme parameterizations has been limited
to data for which the comparison between theory and experiment is
model independent. This is no longer the case for single-particle energies,
for which there exist several conflicting definitions that often even
do not correspond to observables~\cite{Dug12a}. Here, we use here the
eigenvalues of the single-particle Hamiltonian. They provide a lowest-order
approximation to separation energies, which should be corrected for
polarization effects \cite{Ben03a}
and the coupling to vibrations, cf.\ Refs.~\cite{Lit06a,Col10a}. In
Fig.~\ref{fig:spenergy}, single-particle energies are
compared to one-nucleon separation energies to or
from doubly-magic nuclei.

For \nuc{40}{Ca} and \nuc{132}{Sn}, the spectra obtained with the
SLyIII.$xx$ parameterizations are very similar to those of SIII. For
\nuc{208}{Pb},  there are several  differences, in
particular concerning the position of high-$j$ levels.

A rule of thumb predicts that a higher effective mass
gives a more compressed spectrum. We are in a good position to check
this rule. The SLyIII.$xx$ have been constructed using exactly the
same protocol but correspond to four values of the effective mass.
They are a modern version of SIII but share with it many similarities.
Looking to Fig.~\ref{fig:spenergy}, a higher effective mass corresponds
indeed to a more compressed spectrum. However, a change in the effective
mass does not correspond to a simple rescaling of the single-particle
spectra. For neutron holes in \nuc{208}{Pb} or neutron particles in
\nuc{132}{Sn}, the relative distances between levels hardly vary at
all. Also, this rule of thumb is already not valid anymore for a change
in the fitting protocol,
as exemplified by SIII. The differences between the single-particle
spectra obtained with SIII and the SLyIII.$xx$ cannot
be due to the effective mass. The SLy5 results are sometimes
very different. In all cases, the reproduction of the experimental gaps
is rather poor. A more detailed analysis would require
to compute directly one-nucleon separation energies, including
correlations beyond the mean-field which are known to give a sizable
contribution to the two-nucleon separation energies to and from
doubly-magic nuclei~\cite{Ben06a,Ben08b}. We present below in
Sec.~\ref{subsect:transactinides} results for
self-consistent  calculation of binding energies of a few very heavy
odd-$A$ well-deformed nuclei, for which correlations beyond
the mean field can be expected to play a lesser role.

\begin{figure*}[t!]
\centerline{
\includegraphics[width=0.225\linewidth, clip]{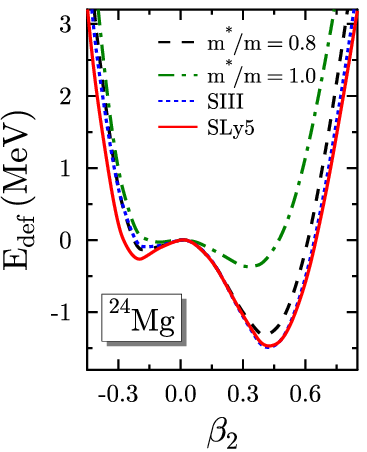}
\includegraphics[width=0.19375\linewidth, clip]{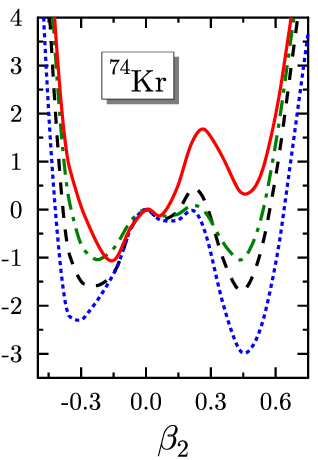}
\includegraphics[width=0.19375\linewidth, clip]{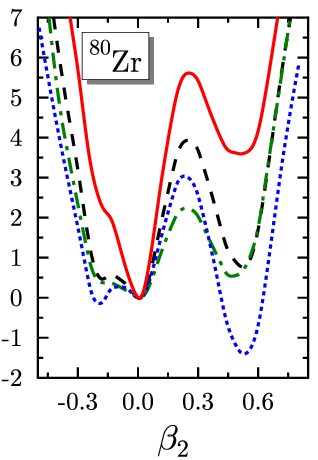}
\includegraphics[width=0.19375\linewidth, clip]{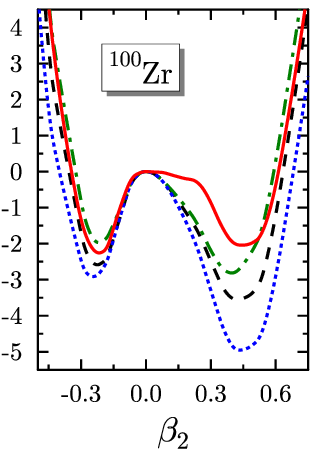}
\includegraphics[width=0.19375\linewidth, clip]{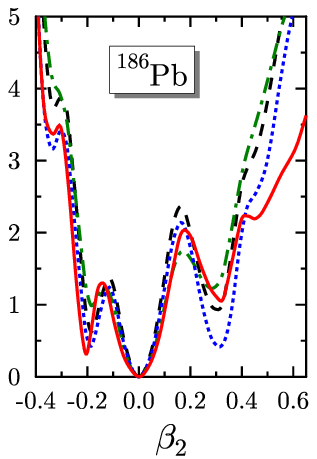}
}
\caption{(Color online)
Potential energy curves along quadrupole deformation $\beta_2$
for $^{24}$Mg, $^{74}$Kr, $^{80}$Zr, $^{100}$Zr, and $^{186}$Pb
obtained by constraint HFB calculations.}
\label{fig:deformation}
\end{figure*}

%
%

\subsection{Deformed nuclei}

In addition to the properties of singly-magic nuclei, we also validate
the performance of the new parameterizations for deformation and rotational
properties of selected key nuclei.

%
%
\subsubsection{Deformation energy curves}

We start by studying nuclei that have been
experimentally identified to be either deformed or have states of
different deformation coexisting at low energy. In Fig.~\ref{fig:deformation},
the deformation energy curves of $^{24}$Mg, $^{74}$Kr, $^{80}$Zr, $^{100}$Zr,
and $^{186}$Pb are plotted as a function of the dimensionless axial
quadrupole deformation
\begin{equation}
\label{eq:beta}
\beta_2
= \sqrt{\frac{5}{16\pi}}\frac{4\pi}{3R^2A} \, \langle 2z^2-x^2-y^2 \rangle
\, ,
\end{equation}
where $R = 1.2 \, A^{1/3}$ fm.
Experimentally, a prolate deformation for their ground state is well
established for $^{24}$Mg and $^{100}$Zr. For $^{80}$Zr, spectroscopic
data suggest that the excited states of the ground state rotational band
have a large quadrupole deformation with a $\beta_{2}$ value around 0.4.
The sparse available data, however, do not rule out that the ground state
of $^{80}$Zr has a complicated structure that involves a large mixing
of different deformations.

The only parameterization that gives rise to a pronounced prolate minimum
for $^{24}$Mg, $^{80}$Zr and $^{100}$Zr is SIII. For both SLyIII.$xx$ shown,
the ground state of $^{80}$Zr is spherical with a prolate minimum at
a slightly higher energy, nearly degenerate with a very shallow oblate
minimum, whereas the ground state of $^{100}$Zr has a
large prolate deformation, with an oblate minimum at smaller
$| \beta_2 |$ excited by around 1~MeV. For SLy5, the ground state of
$^{80}$Zr is spherical with a prolate minimum excited by
about 4~MeV. For $^{100}$Zr, this parameterization gives
nearly degenerate prolate and oblate minima. Before drawing conclusions
on how well these topographies are compatible with experimental data,
one has to estimate how the correlations that we plan to introduce
explicitly in future applications might change the simple picture of
energy curves.
Rodr{\'i}guez and Egido~\cite{Rod11a} have calculated the energy surface
of $^{80}$Zr including triaxial quadrupole deformations using the Gogny
force. They have found several spherical, axial and triaxial minima.
Before projection, the axial part of their energy surface is similar
to the one obtained here with SLy5. For this nucleus, however,
projection on angular momentum alters the topography of the energy
surface, leading after configuration mixing to a ground state with a
predominant component at a large quadrupole axial deformation.

A beyond-mean-field study of the neutron-deficient Kr isotopes using
the Skyrme parameterization SLy6 has been published in Ref.~\cite{Ben06b}.
After projection and mixing,
the relative energy of prolate and oblate states leads
to excitation spectra in disagreement with the experimental data.
The energy curve obtained  for $^{74}$Kr with SLy5 resembles
the one of SLy6 presented, such that it can be
expected that SLy5 would also give similar results
after configuration mixing. By contrast, the prolate minimum obtained with
SIII and SLyIII.\textit{xx}  seems more realistic in view of the experimental
data. Finally, the deformation energy curve of $^{186}$Pb is alike for
all parameterizations, displaying a spherical ground state and a prolate
and oblate minimum within less than 1 MeV excitation energy each.

It is remarkable that SIII and the SLyIII.\textit{xx} parameterizations
give a much more realistic description of the energy curves in the
$A \approx 80$ region than the SLy$x$ parameterizations. This difference,
however, cannot be traced back directly to the linear density dependence,
as some other Skyrme parameterizations with non-integer exponents $\alpha$
of the density dependence give an energy curve for \nuc{80}{Zr} that
is much closer to the one SIII than that of SLy5 \cite{Rei99a}.

The value of the effective mass has a clear effect on the variation
of energy with deformation. Comparing the curves obtained with SLyIII.0.8
and SLyIII.1.0, one can see that a higher effective mass results
in a flatter behavior of the deformation energy curves.

Overall, the SLyIII.\textit{xx} parameterizations provide encouraging
results for the deformation properties at the mean-field level.
The following examples, however, will illustrate some limitations
of these parameterizations.

%
%
\subsubsection{Fission barrier of  $^{240}$Pu}

\begin{figure}[t!]
\includegraphics[width=0.9\linewidth, clip]{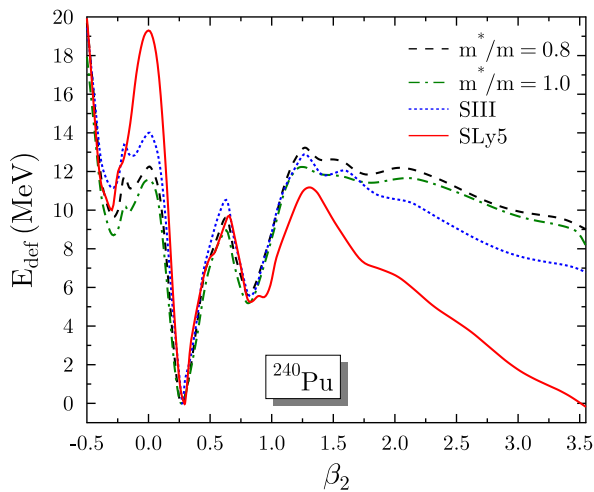}
\caption{(Color online)
Fission barrier along quadrupole deformation $\beta_2$
for $^{240}$Pu obtained by different Skyrme parameter sets.
}\label{fig:fissionbarrier}
\end{figure}
In Fig.~\ref{fig:fissionbarrier}, the fission barrier of $^{240}$Pu
is presented as a function of quadrupole deformation. For all
parameterizations, triaxiality was taken into account in the calculation
of the first barrier and octupole deformations for the second barrier.

In all cases shown in the figure, the excitation energy of the fission
isomer overestimates the experimental value, for which two conflicting
values of $\sim 2.8$ MeV \cite{Sin02a} and $2.25 \pm 0.20$ MeV \cite{Hun01}
can be found in the literature. In the same way, the energies of the inner
and outer fission barriers overestimate the experimental values of 6.05~MeV
and 5.15~MeV respectively \cite{Cap09a}. For SIII this deficiency has been
known for long~\cite{Bar82a}.
Also, the results obtained with SLy5 are less realistic than
those obtained with the SLy4 and SLy6 parameterizations discussed
in Ref.~\cite{Ben04a}. However, one must take into account that the
calculations performed in Ref.~\cite{Ben04a} and here are not fully
equivalent: the pairing strength is not the same and particle number
projection was performed in~\cite{Ben04a} and is not here.
In that paper, it was shown that at the
mean-field level, the energy of the fission isomer is
close to the experimental value with SLy6, whereas SLy4 gives better
agreement when beyond-mean-field correlations are taken into
account. All parameterizations shown in Fig.~\ref{fig:fissionbarrier}
give energies for the fission isomer much larger than SLy4 when used
with standard pairing,
and beyond-mean-field correlations cannot be expected to be
large enough to obtain agreement with the data for any of them.

The differences in barrier height for SIII and SLyIII.\textit{xx}
seen in Fig.~\ref{fig:fissionbarrier} cannot be correlated with the
value of the surface energy coefficient of these parameterizations. The
values for SLy5 ($a_{\text{sym}} = 18.5$~MeV) and SIII (18.6~MeV)
are very similar, whereas those for SLyIII.0.8 (19.5~MeV) and SLyIII.1.0
(19.4~MeV) are significantly larger. The value
of the isoscalar effective mass, and thereby the average level density
at the Fermi energy, does not play a crucial role either.
However, the similarity of the energy curves
obtained with SIII and all SLyIII.\textit{xx} hints at an insufficiency
of a simple linear density dependence to describe large deformation.

%
%
\subsubsection{Superdeformed rotational band in $^{194}$Hg}
\label{subsect:hg194}

\begin{figure}[t!]
\includegraphics[width=0.95\linewidth, clip]{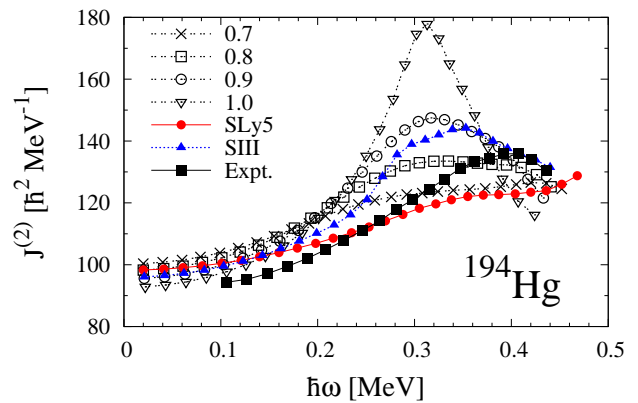}
\caption{\label{fig:sdband}
(Color online)
Dynamical moment of inertia as a function of angular frequency
$\hbar \omega$ for the superdeformed rotational band in $^{194}$Hg
obtained with the parameterizations as indicated.
}
\end{figure}

The next test of the new parameterizations concerns the description of
superdeformed rotational bands (SD). These bands are well described all
over the nuclear chart by self-consistent mean-field calculations and
represent one of the most impressive successes of these
approaches in the 1990's. The SD bands in the Hg region are of specific
interest as the gradual increase of the dynamical moments of inertia
$\mathcal{J}^{(2)}$
\begin{equation}
\label{eq:results:dynmom}
\mathcal{J}^{(2)}
= \frac{\partial \langle J_{z}\rangle}{\partial\omega}
= \frac{1}{\omega} \, \frac{\partial \mathcal{E}}{\partial\omega}
\, ,
\end{equation}
as a function of rotational frequency $\hbar \omega$ results from
the gradual disappearance of pairing correlations and the alignment
of the intruder orbitals. For further details we refer to our recent
detailed analysis of the various contributions of the EDF to
$\mathcal{J}^{(2)}$ in Ref.~\cite{Hel12a}. The dynamical moments of
inertia for the ground SD band of $^{194}$Hg are presented in
Fig.~\ref{fig:sdband}. For SIII and SLyIII.\textit{xx}, the peak in the
$\mathcal{J}^{(2)}$ appears at too low an $\hbar\omega$ and, overall,
the description of the experimental data is less satisfactory than that
of other Skyrme parameterizations such as SLy4 or SkM*. It should
be noted that the currently used pairing strength $V_{0}$ is rather
low in comparison with the typical values of $V_{0}=1250$ MeV fm$^{3}$
that was determined in SD bands. An increase of the pairing strength,
however, will have very little influence on the location of the peak
in $\mathcal{J}^{(2)}$.

%
%
\subsubsection{Single-particle levels in deformed transactinide nuclei}
\label{subsect:transactinides}

\begin{figure}[t!]
\includegraphics[width=0.85\linewidth, clip]{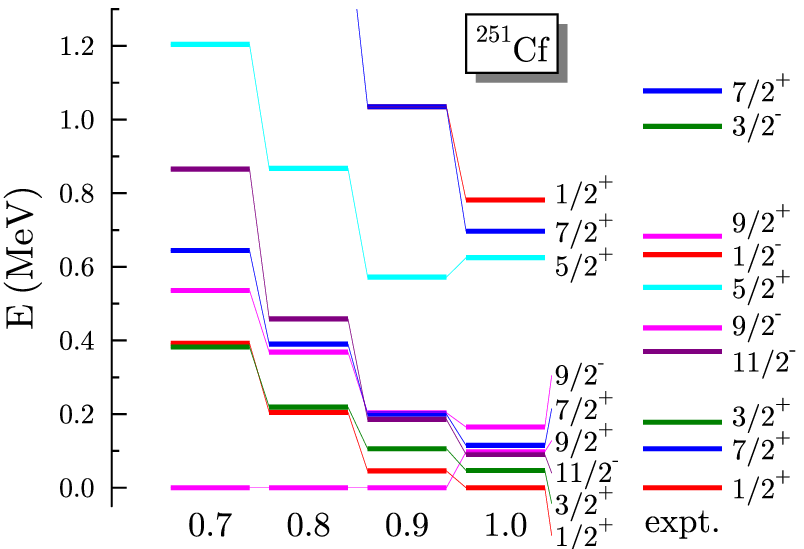}
\includegraphics[width=0.85\linewidth, clip]{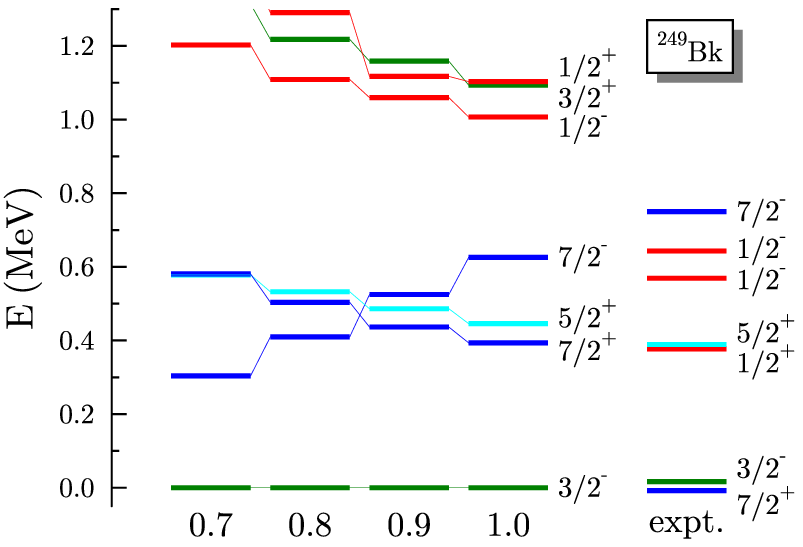}
\caption{\label{fig:transactinide}
(Color online)
Low-lying one-quasiparticle states in $^{251}$Cf and $^{249}$Bk. Experimental data
taken from \cite{Art99,Ahm00}.
}
\end{figure}

Previous studies of odd-mass transactinides~\cite{Ben03b} have put
into evidence some major drawbacks in the spectra obtained with the
current Skyrme parameterization s. We have tested the parameterizations
that we have constructed in this work on two nuclei $^{251}$Cf and
$^{249}$Bk, for which very detailed data are available and which
have been studied in Ref.~\cite{Ben03b}. The same method as in
Ref.~\cite{Ben03b} has been used. Each state results from a
self-consistent calculation of a one-quasiparticle excitation on an even-even
HFB vacuum. In this way, the polarization effect due to the quasiparticle
excitation and the terms in the Skyrme EDF depending on time-odd
densities are taken into account self-consistently.
The results are shown in Fig.~\ref{fig:transactinide}. The $^{251}$Cf
spectra exhibit the expected effect of the effective mass: the
spectrum is becoming more dense when the effective mass is increased.
Note, however, that the compression of the spectrum is not uniform and
that the changes do not correspond to a simple scaling proportional to
the ratios of effective masses as it is sometimes assumed~\cite{Afa12x}.
Moreover, the order of the levels can be different when comparing
the parameterizations. The non-trivial effective mass dependence
is still more apparent for the spectrum of $^{249}$Bk, where the
first excited state is lower in energy for the lowest values of the
effective mass and does not have the same quantum numbers for all the
parameterizations. Although the obtained spectra depend on the
parameterizations, none of the SLy.$xx$ corrects the main drawbacks
of previous EDF parameterizations, {\em i.e.} the misplacement of some levels
that may be connected with specific spherical single-particle orbitals.

%
%
\subsubsection{Particle number symmetry restored deformation energy surfaces}
\label{subsect:mredf}

\begin{figure}[t!]
\includegraphics[width=0.7\linewidth]{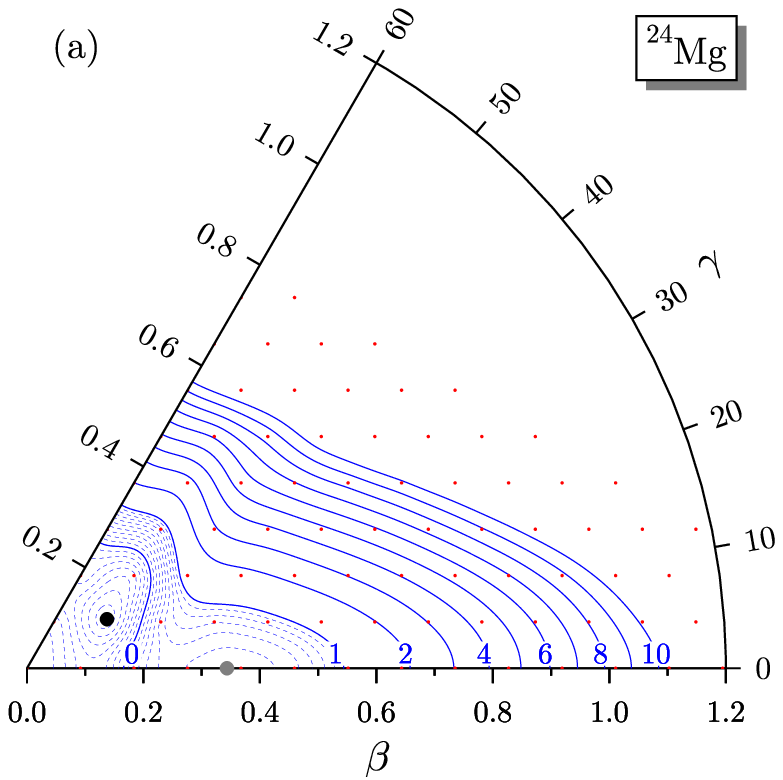}
\includegraphics[width=0.7\linewidth,clip]{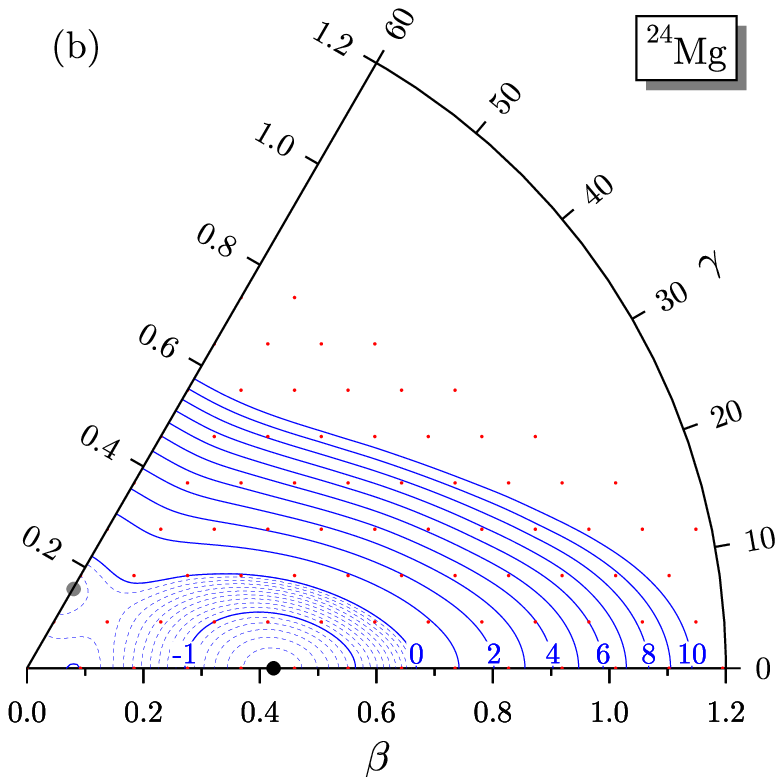}
\caption{\label{fig:mredf}
(Color online)
Particle-number restored deformation energy surface of \nuc{24}{Mg}
in the $\beta$-$\gamma$ plane constructed with the
SLyIII.08 MR EDF without (panel a) and with regularization
(panel b) of the MR EDF. Black filled circles indicate the location
of the absolute minimum of each energy surface, and grey circles the
location of secondary minima. The dots indicate the calculated points.
}
\end{figure}

The main motivation of the present study was to construct a Skyrme
functional that can be used in regularized MR EDF calculations.
In Fig.~\ref{fig:mredf}, we show as an example of such a calculation
the particle-number restored deformation energy surfaces of \nuc{24}{Mg}
without and with regularization in one sextant of the $\beta$--$\gamma$
plane. The calculations were performed as described in Ref.~\cite{Ben09a}
with two differences. The first one is the use of SLyIII.0.8, and
the second one is the use of the extension of the regularization scheme
to trilinear terms in the same particle species as is required by
this parameterization. We use Fomenko's prescription~\cite{Ben09a}
with 19 discretization points for the gauge-space integrals. At small
deformation, the difference between the regularized and non-regularized
energy surfaces is quite dramatic. Without regularization, the absolute
minimum is located in a region where the spurious contribution to the
EDF is particular large. It is only with the regularization that one
finds the usual topography of the energy surface with a prolate axial
minimum.

The nature and size of problems with spurious contributions to the
MR EDF depend strongly on the parameterization of the functional. The
presence of terms that are trilinear in the same particle species
in SLyIII.0.8 makes the deformation and discretization dependence of
the spurious energies much more violent than what is found for the
SIII parameterization used in the regularized calculations Ref.~\cite{Ben09a}.
Also, there are no evident problems in the (non-regularized) particle-number
projected energy surfaces of \nuc{24}{Mg} obtained with SLy4 and presented
in Ref.~\cite{Ben08a}. There, we encountered obvious irregularities only
when projecting simultaneously on particle number and angular momenta
$J > 0$.

A detailed discussion of the regularization that will also address
its application to angular-momentum projection and general configuration
mixing will be given elsewhere~\cite{Ben12R}.

%
%

\section{Summary}
\label{sec:summary}

The present study is a part of our program to construct
an effective interaction of high spectroscopic quality for mean-field
and beyond-mean-field calculations.
In this first step, we have constructed a regularizable (in the sense
of \cite{Lac09a}) EDF within the standard form of the Skyrme EDF.
This requires that the power $\alpha$ of the density dependence
takes an integer value.
The simple form of the non-momentum dependent trilinear terms used here
has known deficiencies. It forbids to obtain a value for
the incompressibility compatible with the empirical value. We have
shown also the problems encountered in the description of charge radii,
fission barriers heights
and moments of inertia in SD bands in the $A \approx 190$ region.
However, the protocol that we have developed leads to a significantly
improved description of shape coexistence in the $A \approx 80$ region.

The four variants with different isoscalar effective mass will
enable studies on how the correlation energy in beyond-mean-field
methods depends on the effective mass. However, the mean-field results
presented here show a clear preference for $m_0^*/m = 0.8$.

Even with their deficiencies, the present parameterizations will
allow us to analyze and benchmark the performance of the regularization.
Work in that direction is underway \cite{Ben12R}

Higher-order terms ({\em i.e.} at least trilinear terms with derivatives)
are clearly necessary to remove the deficiencies of the SLyIII.\textit{xx}
parameterizations pinpointed here and to improve the predictive power
of regularizable Skyrme-type functionals. Work in that direction is also
underway \cite{Sad11t,Sad12x}. Alternative (non-Skyrme-type) regularizable
forms of the density dependence might be considered as well, cf.\ for
example the form proposed in Ref.~\cite{Gez10a}.
The moment the form of a sufficiently flexible functional that is
safely usable in MR EDF calculations has been established, fits should
be performed on the level of MR EDF.

%
%
\section*{Acknowledgments}

This research was supported in parts by
the PAI-P6-23 of the Belgian Office for Scientific Policy, by the F.R.S.-FNRS (Belgium), by the
European Union's Seventh Framework Programme ENSAR under grant
agreement n°262010,
the French Agence Nationale de la Recherche under Grant
No.~ANR 2010 BLANC 0407 "NESQ",
and by the CNRS/IN2P3 through the PICS No.~5994.
Part of the computer time for this study was provided by the computing facilities MCIA (M{\'e}socentre de Calcul Intensif Aquitain) of the Universit{\'e} de Bordeaux and of the Universit{\'e} de Pau et des Pays de l'Adour.

%
%


\begin{thebibliography}{99}

\bibitem{Ben03a}
M. Bender, P.-H. Heenen, and P.-G. Reinhard,
Rev. Mod. Phys. \textbf{75}, 121 (2003).

\bibitem{Sky56a}
T. H. R. Skyrme,
Phil. Mag. \textbf{1}, 1043 (1956).

\bibitem{Sky58a}
T. H. R. Skyrme,
Nucl. Phys. \textbf{9}, 615 (1958).

\bibitem{Vau72a}
D. Vautherin and D. M. Brink,
Phys. Rev. C \textbf{5}, 626 (1972).

\bibitem{Vau73a}
D. Vautherin,
Phys. Rev. C \textbf{7}, 296 (1973).

\bibitem{Bei75a}
M. Beiner, H. Flocard, Nguyen Van Giai, and P. Quentin,
Nucl. Phys. A \textbf{238}, 29 (1975).

\bibitem{Bla76a}
J.-P. Blaizot, D. Gogny, and B. Grammaticos,
Nucl. Phys.  A {\bf 265}  315 (1976).

\bibitem{Bla80a}
J.-P. Blaizot,
Phys. Rep {\bf 64}, 171 (1980).

\bibitem{Col04a}
G. Col{\`o}, Nguyen Van Giai, J. Meyer, K. Bennaceur, and P. Bonche,
Phys. Rev. C {\bf 70}, 024307 (2004).

\bibitem{Cha75a}
B. D. Chang,
Phys. Lett. B {\bf 56}, 205 (1975).

\bibitem{Back75a}
S. O. B{\"a}ckman, A. D. Jackson, and J. Speth,
Phys. Lett. B \textbf{56}, 209 (1975).

\bibitem{War76a}
M. Waroquier, K. Heyde, H. Vinckx,
Phys. Rev. C \textbf{13}, 1664 (1976).

\bibitem{Str76a}
S. Stringari, R. Leonardi, and D. M. Brink,
Nucl. Phys. A {\bf 269}, 87 (1976).

\bibitem{Bla76b}
J. P. Blaizot,
Phys. Lett. B \textbf{60}, 435 (1976).

\bibitem{Bei74a}
M. Beiner and R. J. Lombard,
Ann. Phys. {\bf 86}, 262 (1974).

\bibitem{Kri80a}
H. Krivine, J. Treiner, and O. Bohigas,
Nucl. Phys. A {\bf 336}, 155 (1980).

\bibitem{Bet71a}
H. A. Bethe,
Ann. Rev. Nucl. Sci. {\bf 21}, 93 (1971).

\bibitem{Koh69a}
H. S. K\"ohler and Y. C. Lin,
Nucl. Phys. A {\bf 136}, 35 (1969).

\bibitem{Koh75a}
H. S. K\"ohler,
Phys. Pep. {\bf 18}, 217 (1975).

\bibitem{Koh76a}
H. S. K\"ohler,
Nucl. Phys. A {\bf 258}, 301 (1976).

\bibitem{Gog75a}
D. Gogny,
in Proceedings of the International Conference on Nuclear
Self-Consistent Fields,
International Centre for Theoretical Physics, Trieste,
edited by G. Ripka and M. Porneuf (North-Holland, Amsterdam, 1975), p. 333.

\bibitem{Dec80a}
J. Decharg{\'e} and D. Gogny,
Phys. Rev. C \textbf{21}, 1568 (1980).

\bibitem{Cha97a}
E. Chabanat, P. Bonche, P. Haensel, J. Meyer, and R. Schaeffer,
Nucl. Phys. A {\bf 627}, 710 (1997).

\bibitem{Cha98a}
E. Chabanat, P. Bonche, P. Haensel, J. Meyer, and R. Schaeffer,
Nucl. Phys. A {\bf 635}, 231 (1998).

\bibitem{Fle04a}
P. Fleischer, P. Kl{\"u}pfel, P.-G. Reinhard, J. A. Maruhn,
Phys. Rev. C \textbf{70}, 054321 (2004).

\bibitem{Pro04a}
L. Prochniak, P. Quentin, D. Sams{\oe}n, and J. Libert,
Nucl. Phys. A \textbf{730}, 59 (2004).

\bibitem{Hee93a}
P.-H. Heenen, P. Bonche, J. Dobaczewski, H. Flocard,
Nucl. Phys. A \textbf{561}, 367 (1993).

\bibitem{Zdu07a}
H. Zdu{\'n}czuk, W. Satu{\l}a, J. Dobaczewski, M. Kosmulski,
Phys. Rev. C \textbf{76}, 044304 (2007).

\bibitem{Ben06a}
M. Bender, G. F. Bertsch, and P.-H. Heenen,
Phys. Rev. C {\bf 73}, 034322 (2006).

\bibitem{Flo76a}
H. Flocard, D. Vautherin
Nucl. Phys. A \textbf{264}, 197 (1976).

\bibitem{BB69a}
R. Balian and E. Brezin,Nuovo Cim. B \textbf{64} 37 (1969).

\bibitem{Bon90a}
P. Bonche, J. Dobaczewski, H. Flocard, P.-H. Heenen and J. Meyer
Nucl. Phys. A {\bf 510}, 466 (1990).

\bibitem{Taj92a}
N. Tajima, H. Flocard, P. Bonche, J. Dobaczewski, and P.-H. Heenen,
Nucl. Phys. A \textbf{542}, 355 (1992).

\bibitem{Ben06b}
M. Bender, P. Bonche, and P.-H. Heenen,
Phys. Rev. C {\bf 74}, 024312 (2006).

\bibitem{Ben08a}
M. Bender and P.-H. Heenen,
Phys. Rev. C {\bf 78}, 024309 (2008).

\bibitem{Les07a}
T. Lesinski, M. Bender, K. Bennaceur, T. Duguet, and J. Meyer,
Phys. Rev. C {\bf 76}, 014312 (2007).

\bibitem{Ben09b}
M. Bender, K. Bennaceur, T. Duguet, P.-H. Heenen, T. Lesinski and J. Meyer,
Phys. Rev. C {\bf 80}, 064302 (2009).

\bibitem{Kor08a}
M. Kortelainen, J. Dobaczewski, K. Mizuyama, J. Toivanen,
Phys. Rev. C \textbf{77}, 064307 (2008).

\bibitem{Rod02a}
R. Rodr{\'i}guez-Guzm{\'a}n, J. L. Egido and L. M. Robledo,
Nucl. Phys. A \textbf{709}, 201 (2002).

\bibitem{Yao10a}
J. M. Yao, J. Meng, P. Ring, and D. Vretenar,
Phys. Rev. C {\bf 81}, 044311 (2010).

\bibitem{Ang01a}
M. Anguiano, J. L. Egido, and L. M. Robledo,
Nucl. Phys. A {\bf 696}, 467 (2001).

\bibitem{Dob07a}
J. Dobaczewski, M. V. Stoitsov, W. Nazarewicz, and P. G. Reinhard,
Phys. Rev. C {\bf 76}, 054315 (2007).

\bibitem{Lac09a}
D. Lacroix, T. Duguet, and M. Bender,
Phys. Rev. C {\bf 79}, 044318 (2009).

\bibitem{Dug09a}
T. Duguet, M. Bender, K. Bennaceur, D. Lacroix, and T. Lesinski,
Phys. Rev. C {\bf 79}, 044320 (2009).

\bibitem{Egi91a}
J. L. Egido and L. M. Robledo,
Nucl. Phys. A \textbf{524}, 65 (1991).

\bibitem{Dug03a}
T. Duguet and P. Bonche,
Phys. Rev. C \textbf{67}, 054308 (2003).

\bibitem{Rob07a}
L. M. Robledo,
Int. J. Mod. Phys. E \textbf{16}, 337 (2007).

\bibitem{Rob10a}
L. M. Robledo,
J. Phys. G  \textbf{37}, 064020 (2010).

\bibitem{Ben09a}
M. Bender, T. Duguet, and D. Lacroix,
Phys. Rev. C {\bf 79}, 044319 (2009).

\bibitem{Sad11t}
J. Sadoudi,
Th{\`e}se, Universit{\'e} Paris-Sud XI, 2011.

\bibitem{Sad12x}
J. Sadoudi, K. Bennaceur, T. Duguet, J. Meyer, \emph{et al.},
\emph{in preparation}.

\bibitem{Kor10a}
M. Kortelainen, T. Lesinski, J. More, W. Nazarewicz, J. Sarich,
N. Schunck, M. V. Stoitsov, and S. Wild,
Phys. Rev C {\bf 82}, 024313 (2010).

\bibitem{Kor11a}
M. Kortelainen, J. McDonnell, W. Nazarewicz, P.-G. Reinhard,
J. Sarich, N. Schunck, M. V. Stoitsov, and S. M. Wild,
Phys. Rev C {\bf 85}, 024304 (2012).

\bibitem{Eng11a}
E. Engel and R. M. Dreizler,
\emph{Density Functional Theory},
(Springer, Heidelberg, Dordrecht, London, New York, 2011).

\bibitem{Hel12a}
V. Hellemans, P.-H. Heenen, M. Bender,
Phys. Rev. C \textbf{85}, 014326 (2012).

\bibitem{Ter95a}
J. Terasaki, P.-H. Heenen, P. Bonche, J. Dobaczewski, and H. Flocard,
Nucl. Phys. \textbf{1593}, 1 (1995).

\bibitem{Kri90a}
S. J. Krieger, P. Bonche, H. Flocard, P. Quentin, and M. S. Weiss,
Nucl. Phys. A {\bf 517}, 275 (1990).

\bibitem{Bro98a}
B. A. Brown,
Phys. Rev. C \textbf{58}, 220 (1998).

\bibitem{Bro00a}
B. A. Brown, W. A. Richter, and R. Lindsay,
Phys. Lett. B \textbf{483}, 49 (2000).

\bibitem{Gor08a}
S. Goriely and J. M. Pearson,
Phys. Rev. C \textbf{77}, 031301(R), (2008).

\bibitem{Ben00b}
M. Bender, K. Rutz, P.-G. Reinhard, and J. A. Maruhn,
Eur. Phys. J. A {\bf 7}, 467 (2000).

\bibitem{Wir88a}
R. B. Wiringa, V. Fiks, and A. Rabrocini,
Phys. Rev. C {\bf 38}, 1010 (1988).

\bibitem{Les06a}
T. Lesinski, K. Bennaceur, T. Duguet, and J. Meyer,
Phys. Rev. C {\bf 74}, 044315 (2006).

\bibitem{Aud03a}
G. Audi, A. H. Wapstra, and C. Thibault,
Nucl. Phys. A {\bf 729}, 337 (2003).

\bibitem{Nad94a}
E. G. Nadjakov, K. P. Marinova, and Y. P. Gangrsky,
At. Data Nucl. Data Tables {\bf 56}, 133 (1994).

\bibitem{Bla05a}
F. Le Blanc \emph{et al.},
Phys. Rev. C {\bf 72}, 034305 (2005).

\bibitem{Ben00a}
M. Bender, K. Rutz, P.-G. Reinhard, and J. A. Maruhn,
Eur. Phys. J. A {\bf 8}, 59 (2000).

\bibitem{Yam09a}
M. Yamagami, Y. R. Shimizu, T. Nakatsukasa,
Phys. Rev. C {\bf 80}, 064301 (2009).

\bibitem{dataset04}
http://orph02.phy.ornl.gov/workshops/lacm08/UNEDF/DataSet04.dat

\bibitem{Dug12a}
T. Duguet and G. Hagen,
Phys. Rev. C \textbf{85}, 034330 (2012) and references therein.

\bibitem{Lit06a}
E. Litvinova and P. Ring,
Phys. Rev. C \textbf{73}, 044328 (2006).

\bibitem{Col10a}
G. Col{\`o}, H. Sagawa, and P. F. Bortignon,
Phys. Rev. C \textbf{82}, 064307 (2010).

\bibitem{Ben08b}
M. Bender, G. F. Bertsch, and P.-H. Heenen,
Phys. Rev. C \textbf{78}, 054312 (2008).

\bibitem{Rod11a}
T. R. Rodriguez and J. L. Egido,
Phys. Lett. B {\bf 705}, 255 (2011).

\bibitem{Rei99a}
P.-G. Reinhard, D. J. Dean, W. Nazarewicz, J. Dobaczewski,
J. A. Maruhn, and M. R. Strayer,
Phys. Rev. C \textbf{60}, 014316 (1999).

\bibitem{Sin02a}
  B. Singh, R. Zywina, and R. B. Firestone,
  Nucl. Data Sheets \textbf{97}, 241 (2002).

\bibitem{Hun01}
  M. Hunyadi, D. Gassmann, A. Krasznahorkay, D. Habs, P. G. Thirolf,
  M. Csatl{\'o}s, Y. Eisermann, T. Faestermann, G. Graw, J. Gulyas,
  R. Hertenberger, H. J. Maier, Z. M{\'a}t{\'a}a, A. Metz,
  and M. J. Chromik,
  Phys. Lett. \textbf{B505}, 27 (2001).

\bibitem{Cap09a}
R. Capote, M. Herman, P. Oblo\v{z}insk{\'y}, P. G.Young, S. Goriely, T.
Belgya, A. V. Ignatyuk, A. J.Koning, S. Hilaire, V. A. Plujko,
M. Avrigeanu, O. Bersillon, M. B. Chadwick, T. Fukahori, Zhigang
Ge, Yinlu Han, S. Kailas, J. Kopecky, V. M. Maslov, G. Reffo,
M. Sin, E. Sh. Soukhovitskii, and P. Talou,
Nucl. Data Sheets \textbf{110}, 3107 (2009);
Reference Input Parameter Library (RIPL-3)
[http://www-nds.iaea.org/RIPL-3/].

\bibitem{Bar82a}
J. Bartel, P. Quentin, M. Brack, C. Guet, and H.-B. H{\aa}kansson,
Nucl. Phys. A \textbf{386}, 79 (1982).

\bibitem{Ben04a}
M. Bender, P.-H. Heenen, and P. Bonche,
Phys. Rev. C {\bf 70}, 054304 (2004).

\bibitem{Jod12x}
R. Jodon, K. Bennaceur, M. Bender, J. Meyer,
\emph{in preparation}.

\bibitem{Ben03b}
M. Bender, P. Bonche, T. Duguet and P.-H. Heenen,
Nucl. Phys. A  \textbf{723}, 354 (2003).

\bibitem{Art99}
A. Artna-Cohen,
Nucl. Data Sheets \textbf{88}, 155 (1999).

\bibitem{Ahm00}
I. Ahmad, M. P. Carpenter, R. R. Chasman, J. P. Greene, R. V. F. Janssens, T. L. Khoo, F. G. Kondev, T. Lauritsen, C. J. Lister, P. Reiter, D. Seweryniak, A. Sonzogni, J. Uusitalo, and I. Wiedenh\"over,
Phys. Rev. C \textbf{62}, 064302 (2000).

\bibitem{Afa12x}
A.V. Afanasjev and S. Shawaqfeh,
Phys. Lett. B \textbf{706}, 177 (2011).

\bibitem{Ben12R}
M. Bender, B. Avez, B. Bally, T. Duguet, P.-H. Heenen, D. Lacroix,
\emph{in preparation}.

\bibitem{Gez10a}
A. Gezerlis and G. F. Bertsch,
Phys. Rev. Lett. \textbf{105}, 212501 (2010).

\end{thebibliography}
\end{document}